\documentclass[11pt,a4paper]{article}
\usepackage[latin1]{inputenc}
\usepackage{multirow}
\usepackage[table,xcdraw]{xcolor}
\usepackage{rotating}
\usepackage{amsmath}
\usepackage{amsfonts}
\usepackage{amssymb}
\usepackage{graphicx}
\usepackage{physics}
\usepackage[export]{adjustbox}
\usepackage{soul} 
\usepackage{tikz} 
\usepackage{tabularx}

\usepackage{cuted}

\usepackage{subcaption}
\usepackage[left=2.00cm, right=2.00cm, top=2.00cm, bottom=2.00cm]{geometry}
\usepackage[sort&compress,square,comma,numbers]{natbib}

\begin{document}
	
\title{Characterization of the 2D Su-Schrieffer-Heeger Model with Second-Nearest-Neighbor Interactions}
\date{25. June 2023}
\maketitle

\author{Chani Stella van Niekerk$^1$, Robert Warmbier$^2$\\
$^1$ Department of Physics, University of Johannesburg, Johannesburg, South Africa\\
Email Address: csvanniekerk@uj.ac.za\\
$^2$ School of Physics, University of the Witwatersrand, Johannesburg, South Africa\\
email Address: robert.warmbier@wits.ac.za}

\begin{abstract}
It is known that a two dimensional dimerized Su-Schrieffer-Heeger model can produce a non-trivial topological phase. It is a simple nearest-neighbor model with either two or four lattice sites in in two dimensions. Su-Schrieffer-Heeger model is easy to analyse but neglects important interaction in physical systems. In this work, an extended version of this model is proposed which includes all possible second nearest neighbor interactions in order to make it more feasible to describe realistic systems. The topological phases and properties of the model are characterized using a polarization invariant. It is further shown that second nearest neighbor interactions can be used to evoke a topological phase transition as well.
\end{abstract}

\section{Introduction}
The discovery of the quantum Hall effect in the 1980's led to the definition of a quantum state where the bulk of a sample (2D at the time) was insulating and the electric current was carried only along the edge of the sample. With the use of the concept of topology, this state could be defined as topologically distinct from all other states of matter known before this. \cite{Zhang}
\\
Materials presenting this state were categorized as topological insulators and have been studied extensively since their discovery. \cite{Zhang, Hasan, hasan2010colloquium, Bernevig, Moore, Mele} Topological insulators have a bulk band gap like normal insulators, however their surface/edge states are conducting. The conducting states are by definition robust and thus resistant to defects and impurities provided the symmetry of the system is preserved.\cite{Hasan} 
\\
When a $d$-dimensional material is in a non-trivial topological state, it can host $(d - n)$ dimensional boundary states where $n = 1, \ldots, d-1$ when it exhibits chiral symmetry. For example, a 2D topological insulator can host 1D (edge) and 0D (corner) conducting states. \cite{Chen} When $n > 1$, the topological insulator falls into the category of higher order topological insulators (HOTIs). HOTIs are often weak topological insulators and have non-trivial phases that generally break down when disorder is added to them and evokes a symmetry change. Topological insulators are strong topological insulators when the surface states are topologically protected from localization, and they are weak otherwise. For example, in a 2D system edge conducting states would indicate a strong topological insulator because the entire surface was conducting, while if the topological states were only in the corners, this would be considered to be a weak topological state. \cite{Arxiv} 
\\
In this work we characterize the non-trivial topological phase of a 2D extended Su-Schrieffer-Heeger (SSH) model with the addition of second nearest neighbor (SNN) interactions. The extended SSH model without SNN interactions has been studied extensively, in particular, it has been shown that this model exhibits both corner and edge states which are robust and non-trivial. \cite{Chen, Arxiv, Liu, Liu1, Obana, Xie, Ota, Zhu, Xu, Bena, Kim, Boma} This forms a solid basis to build upon. To our knowledge next second neighbor interactions have not been studied fully yet, literature only focuses on one type of interaction for example, in the work done by Xu \textit{et al.} \cite{Xu}, only intra-cellular interactions are considered. We consider all SNN interactions and their effects on the topological properties of the system.

\section{Theoretical Model}
\begin{figure}[htb]
	\centering
	\includegraphics[width=8.6cm]{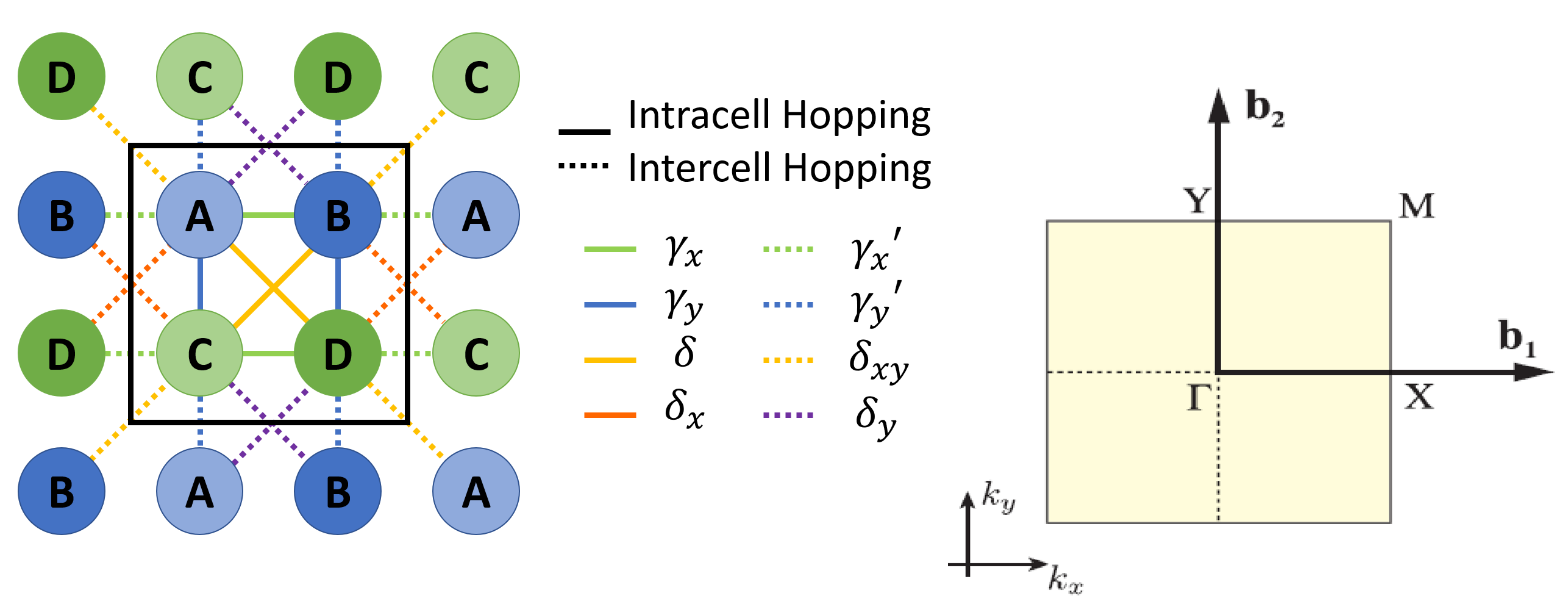}
	\caption{Construction of a 2D system using two different 1D SSH chains coupled together in the $y$-direction. Lattice sites A, B, C, and D represent one unit cell (outlined in black). Dashed and solid lines show all the possible nearest and next nearest neighbor hoppings from each unit cell.}
	\label{Fig: mod_diag}
\end{figure}

In this paper, the extended SSH model is comprised of a set of interacting 1D SSH arrays coupled together to form a 2D system as shown in Figure \ref{Fig: mod_diag}. Each 1D array consists of dimerized lattices along the $x$-direction each with two sublattice sites separated by a distance $d$. The 2D system is built from coupling the chains in a similar way along the $y$-direction. In order to keep the system general, two different 1D chains are used - one with sublattice sites A and B and another with sublattice sites C and D resulting in a ABCD lattice site unit cell. \\

\subsection{Real Space Model}
Using tight-binding principles, as with a 1D SSH model, the 2D system can be split into its nearest neighbor (NN) and second nearest neighbor (SNN) interactions. The resultant real space Hamiltonian $H = H_{NN} + H_{SNN}$ consists of
\begin{equation}	\label{Equ: Real H}
	\begin{split}
		H_{NN} = \sum_{n, m} \Bigg\{  &
		\bigg[\gamma_{x} \ket{A_{n, m}} + \gamma_x^\prime \ket{A_{n+1, m}}\bigg] \bra{B_{n,m}} +  \\
		& \bigg[\gamma_{x} \ket{C_{n, m}} + \gamma_x^\prime \ket{C_{n+1, m}}\bigg] \bra{D_{n,m}} +	\\
		& \bigg[\gamma_y \ket{A_{n, m}} + \gamma_y^\prime \ket{A_{n, m+1}}\bigg]\bra{C_{n, m}}  + \\ & \bigg[\gamma_y \ket{B_{n, m}} + \gamma_y^\prime \ket{B_{n, m+1}}\bigg]\bra{D_{n, m}} + \mathrm{h.c.} \Bigg\},
	\end{split}
\end{equation}
and
\begin{equation}
	\begin{split}
		H_{SNN} = \sum_{n, m} \Bigg\{&
		\bigg[\delta \ket{A_{n, m}} + \delta_x \ket{A_{n-1, m}} + \delta_y \ket{A_{n, m-1}} + \\ & \delta_{xy} \ket{A_{n-1, m-1}}\bigg]\bra{D_{n,m}} + \\
		& \bigg[\delta\ket{B_{n, m}} + \delta_x \ket{B_{n-1, m}} + \delta_y \ket{B_{n, m+1}} + \\ & \delta_{xy} \ket{B_{n-1, m+1}}\bigg] \bra{C_{n,m}} + \mathrm{h.c.} \Bigg\}.
	\end{split}.
\end{equation}
Here $n, m$ refer to the index of the unit cells in the $x$ and $y$ directions, $\gamma_{l}$ and $\gamma_{l}^\prime$; $l \in \{x, y\}$ are the nearest intra-cell and inter-cell interactions, and $\delta$ and $\delta_{p}^\prime, \;\; (p \in \{x, y, xy\})$, are the intra-cell and inter-cell second nearest neighbor interactions. When $p = \{x, y\}$, the inter-cell hopping is across the edge of the unit cell and when $p = xy$, the hopping crosses the corner of the cell. 
\\
The real space system is studied using a finite number of repeated unit cells assuming both periodic and non-periodic boundary conditions. Periodic boundary conditions are simulated by connecting the first and last cells in the chosen periodic direction, while non-periodic boundary conditions are applied by leaving the first and last cells as free ends. In this work we only study the non-periodic and completely periodic cases, the case with mixed boundary conditions follows trivially from those.
\\
For non-periodic systems, the allowed states can be split into bulk, edge and corner states when chiral symmetry is present. These states are defined on the localization of their probability density distribution. Similarly, when chiral symmetry is not present, surfaces states localized on the corner and edge or a combination of the two are defined in the same way. This categorization allows the confirmation of the topological states which occur in the system. Surface planes can be chosen to align with the edges of the unit cell or cut through the center of the cell. For bipartite lattices these are not equivalent. Cutting through the centre of the cell is equivalent to the shifts $\gamma_x \rightleftarrows \gamma_x^\prime$, $\gamma_y \rightleftarrows \gamma_y^\prime$, $\delta \rightleftarrows \delta_{xy}$, and $\delta_x \rightleftarrows \delta_y$. As this would belong to different topological states, as will be seen, the latter case is not considered here.

\subsection{Reciprocal Space Model}
\label{Section: k Model}
Hermitian systems have a direct relation between the topological properties of the bulk reciprocal space system and the surface states in real space due to bulk boundary correspondence. By assuming the system is infinitely periodic and Fourier transforming the real space Hamiltonian, Bloch wave solutions of the form
\begin{equation}
	\ket{\psi(\va{k})} = \sum_{X} \sum_{n, m} \left[ e^{i \va{k} \cdot (n, m)} \phi_X(\va{k}) \ket{X_{n, m}}\right]
\end{equation}
with $ X\; \in \; \{A, B, C, D\}$ are found. Since $E \phi = h(\va{k}) \phi$, the corresponding reciprocal space Hamiltonian in the basis $\hat \psi_{\va k}^\dagger = ( \hat A,  \hat B,  \hat C,  \hat D)$ can be written as:
\begin{equation}
	H = \hat h_{SSH}(\va k) = \mqty( 0 & s_x(k_x) & s_y(k_y) & s_{xy}(\va{k}) \\
	s^*_x (k_x) & 0 & s_{xy}^\prime(\va{k}) & s_y(k_y) \\
	s_y^{*} (k_y) & s_{xy}^{\prime *}(\va{k}) & 0 & s_x(k_x) \\
	s_{xy}^*(\va{k})& s_y^{*} (k_y) & s_x^* (k_x) & 0
	),
\end{equation}
where $s_l (k_l) = \gamma_l + \gamma_l^\prime e^{-ik_l}, \; l \in \{x, y\}$ and $s_{xy}(\va{k}) = \delta + \delta_x e^{-ik_x} + \delta_y e^{-ik_y} + \delta_{xy} e^{-i(k_x + k_y)}$ and $s_{xy}^\prime(\va{k}) =  \delta +  \delta_x e^{ik_x} + \delta_y e^{-ik_y} + \delta_{xy} e^{-i(-k_x + k_y)}$. 

\section{Properties of the model}
\label{Section: Prop of model}
Due to bulk boundary correspondence, by studying the reciprocal space Hamiltonian, one can derive the topological properties of the overall system. In particular, one can establish for which regions of the parameter space the system will be topological. These properties are protected by a set of symmetries defined using the Hamiltonian. 
\\
In order for the topological properties of an insulating system to change, a band gap closure or band inversion needs to occur. The closure is a necessary, but insufficient, condition for the transition. To confirm that the topological phase changes, an invariant quantity derived from the Hamiltonian must accompany the gap closure. 
\\
Band inversions and invariant quantities can be defined for metallic systems as well, however, due to the number of band interactions at the Fermi level, quantifying the change in phase is challenging and beyond the scope of this work.  

\subsection{Energy Eigenvalues}
\label{Section: Eigs of model}
Gap closures act as an initial marker for a potential topological phase transition. Degeneracy in the bands can be found by setting the discriminant of the characteristic polynomial describing the energy eigenvalues to zero. The gap closure conditions describe the points in parameter space where the degeneracy occurs, but do not distinguish why a degeneracy occurs making it a necessary but not sufficient condition for a phase transition. 
\\
When SNN interactions are excluded, the energy values are given by 
\begin{equation}
	E = \pm \;\Big|\;|s_x| \pm |s_y|\;\Big|.
\end{equation}
In order for the valence and conduction bands to have the same energy, $\frac{\gamma_{l}^\prime}{\gamma_{l}} = -e^{\pm ik_l}$, with $l \in \{x,y\}$, $\bigg|\frac{\gamma_{x}^{(\prime)}}{\gamma_{y}^{(\prime)}}\bigg|=1$,   $\bigg|\frac{\gamma_{x(y)}^{\prime}}{\gamma_{y(x)}}\bigg|=1$, or $\frac{\gamma_{x} \gamma_{x}^\prime}{\gamma_{y} \gamma_{y}^\prime}=\frac{\cos(k_y)}{\cos(k_x)}$. These conditions hold in parameter space for at least one $k$-value. For example, the conditions simplify to $\frac{\gamma_{l}^\prime}{\gamma_{l}} = \mp 1$ for $k_l = 0, \pi$ and $\frac{\gamma_{x} \gamma_{x}^\prime}{\gamma_{y} \gamma_{y}^\prime}=\pm 1$ when $k_x = k_y = 0$ or $\pi$ or $k_{x(y)} = 0$ and $k_{y(x)} = \pi$.
\\
The addition of SNN interactions gives a more complicated system of energy eigenvalues: 
\\
\begin{equation}
	\begin{aligned}
	\label{Equation: Full Energy Diag}
	E_{-\pm} & = -\frac{\sqrt{3}}{6} \left[\sqrt{F_2+2 A_2} \pm \sqrt{-F_2+4
		A_2-\frac{12 \sqrt{3} C_3}{\sqrt{F_2+2 A_2}}}\right] 
	\\
	E_{+\pm} & = \frac{\sqrt{3}}{6} \left[\sqrt{F_2+2 A_2} \pm \sqrt{-F_2+4
		A_2+\frac{12 \sqrt{3} C_3}{\sqrt{F_2+2 A_2}}}\right]
	\end{aligned}
\end{equation}
with 
\begin{equation}
	\begin{aligned}
	A_2 & = 2 \left(|s_x|^2 + |s_y|^2\right) + |s_{xy}|^2 + |s_{xy}^{\prime}|^2 \\
		B_4 & = \left(|s_x|^2 - |s_y|^2\right)^2  \\ &	+ |s_{xy}|^2|s_{xy}^\prime|^2
-2\mathrm{\;Re}\left(s_{xy}^*  \left(s_x^2 s_{xy}^\prime + s_y^2 s_{xy}^{\prime *} \right)\right)\\ 
C_3 & = \mathrm{\;Re}\left(s_x \left(s_{xy}^\prime s_y^* + s_y s_{xy}^*\right)\right) \\
D_6 & = \sqrt{- (A_2^2 + 12 B_4)^3 +  \left(A_2^3 - 36 A_2 B_4 - 54 C_3^2\right)^2} \\ & - \left(A_2^3 - 36 A_2 B_4 - 54 C_3^2\right) \\
F_2 & = \frac{ \left(A_2^2+12 B_4\right)}{\sqrt[3]{D_6}} + \sqrt[3]{D_6}.
	\end{aligned}
\end{equation}

For this model meaningful analytical discriminant conditions cannot not be given for the general case.

\subsection{Wave Functions}
In order to fully define the topological phase transition, a gap closure needs to be accompanied by a change in an invariant quantity. Invariants are defined using the wave functions of the model Hamiltonian. For the general 2D SSH model without SNN interactions, the analytic wave function is expressed as follows:
\begin{equation}
\begin{aligned}
	\Psi_{-} = \left\{-\frac{\sqrt{s_x s_y}}{2 \sqrt{{s_x}^*{s_y}^*}}
,+(-)\frac{\sqrt{s_y} \;\;\Big| \; |s_x| - |s_y|\; \Big|}{2\sqrt{s_y^*}\left(|s_x| - |s_y|\right)}, \right.\\
 \left. -(+)\frac{\sqrt{s_x} \;\;\Big| \; |s_x| - |s_y|\; \Big|}{2\sqrt{s_x^*}\left( |s_x| - |s_y|\right)},\frac{1}{2}\right\} \\
\Psi_{+} = \left\{\frac{\sqrt{s_x s_y}}{2 \sqrt{{s_x}^*{s_y}^*}}
,-(+)\frac{\sqrt{s_y} \;\;\Big| \; |s_x| + |s_y|\; \Big|}{2\sqrt{s_y^*}\left(|s_x| + |s_y|\right)}, \right. \\
\left. -(+)\frac{\sqrt{s_x}\;\;\Big| \; |s_x| + |s_y|\; \Big|}{2\sqrt{s_x^*}\left( |s_x| + |s_y|\right)},\frac{1}{2}\right\}
\end{aligned}
\end{equation}

As with the energy expressions, adding SNN interactions makes the expressions more complicated:
\begin{equation}
	\begin{aligned}
	\Psi_{-\pm} & =	\frac{\left\{\frac{r_{1\pm}}{q_{1\pm}},\frac{n_{1\pm}}{q_{1\pm} },-\frac{s_{1\pm}}{q_{1\pm}},1\right\}}{ \sqrt{\left| \frac{n_{1\pm}}{q_{\pm}}\right| ^2+\left|
		\frac{r_{1\pm}}{q_{1\pm}}\right| ^2+\left| \frac{s_{1\pm}}{q_{1\pm}}\right|
		^2+1}} \\
\Psi_{+\pm} & = 		\frac{\left\{\frac{r_{2\pm}}{q_{2\pm}},\frac{n_{2\pm}}{q_{2\pm} },-\frac{s_{2\pm}}{q_{2\pm}},1\right\}}{ \sqrt{\left| \frac{n_{1\pm}}{q_{\pm}}\right| ^2+\left|
		\frac{r_{2\pm}}{q_{2\pm}}\right| ^2+\left| \frac{s_{2\pm}}{q_{2\pm}}\right|
		^2+1}}
	\end{aligned}
\end{equation}
with 
\begin{equation*}
	\begin{aligned}
g &= \sqrt[3]{2 A^2 + F^2}\\
k^2 &= \frac{4A^2-F^2}{3}\\
l^2 &= \frac{4\sqrt{3}C^3}{\sqrt{2A^2+F^2}} \\
m_{1\pm} & = g \pm 3 \sqrt{(k-l) (k+l)}\\
m_{2\pm} & = - g \pm  3 \sqrt{l^2+k^2} \\
n_{1(2)\pm} & = -\frac{m_{1(2)\pm} s_y s_x^*}{3}+s_{xy} \left(s_x^*\right)^2-|s_{xy}|^2
{s_{xy}^\prime} \\ &+s_y^2 s_{xy}^*+\frac{m_{1(2)\pm}^2 {s_{xy}^\prime}}{36} \\
q_{1(2)\pm} & =s_x^* \left(\left| s_y\right| ^2+\frac{m_{1(2)\pm}^2}{36}\right)+s_{xy}^*
\left(s_x {s_{xy}^\prime}-\frac{m_{1(2)\pm} s_y}{6}\right) \\ &-\frac{m_{1(2)\pm} {s_{xy}^\prime}
	s_y^*}{6}-s_x \left(s_x^*\right)^2\\
r_{1(2)\pm} & = -s_y \left| s_y\right| ^2 +|s_x|^2 s_y+s_{xy}
{s_{xy}^\prime} s_y^* \\ &+\frac{m_{1(2)\pm}}{36} \left(-6 s_{xy} s_x^*+m_{1(2)\pm}
s_y-6 s_x {s_{xy}^\prime}\right)\\
s_{1(2)\pm} & = \frac{1}{216} \left(m_{1(2)\pm}^3-36 m_{1(2)\pm} |s_y|^2\right) \\ & +s_{xy}^* \left(s_x
s_y-\frac{m_{1(2)\pm} s_{xy}}{6}\right)+s_x^* \left(s_{xy} s_y^*-\frac{m_{1(2)\pm}
	s_x}{6}\right).
	\end{aligned}
\end{equation*}

\subsection{Symmetry}
\label{Section: Symm of model}
In addition to the wave function, the symmetry of the system must be defined in order to choose an appropriate invariant. While several symmetries can be defined for a system, choosing an appropriate invariant hinges on time-reversal ($TH(\va{k})T^{-1} = H(-\va{k})$), particle-hole($PH(\va{k})P^{-1} = -H(-\va{k})$), chiral($CH(\va{k})C^{-1} = -H(\va{k})$), and inversion ($\iota H(\va{k}) \iota^{-1} = H(- \va{k})$) symmetries. 
For a spinless system, these symmetries are given by 
\begin{equation*}
	\mathrm{Time\; reversal:} \;\; T = \kappa
\end{equation*}
\begin{equation*}
	\mathrm{Particle \;Hole:} \;\; P = \sigma_z^{(x)} \otimes \sigma_z^{(y)}\kappa =\begin{pmatrix}
		1 & 0 & 0 & 0 
		\\
		0 & -1 & 0 & 0
		\\
		0 & 0 & -1 & 0
		\\
		0 & 0 & 0 & 1
	\end{pmatrix} \kappa
\end{equation*}
\begin{equation*}
	\mathrm{Chiral:}\;\; C = \sigma_z^{(x)} \otimes \sigma_z^{(y)} =\begin{pmatrix}
		1 & 0 & 0 & 0 
		\\
		0 & -1 & 0 & 0
		\\
		0 & 0 & -1 & 0
		\\
		0 & 0 & 0 & 1
	\end{pmatrix} 
\end{equation*}
\begin{equation*}
	\mathrm{Inversion:}\;\; \iota = \sigma_x^{(x)} \otimes \sigma_x^{(y)}  =\begin{pmatrix}
		0 & 0 & 0 & 1 
		\\
		0 & 0 & 1 & 0
		\\
		0 & 1 & 0 & 0
		\\
		1 & 0 & 0 & 0
	\end{pmatrix}
\end{equation*}
where $\sigma_x^{(l)}$ and $\sigma_z^{(l)}$ correspond to Pauli matrices acting in the sublattice subspace in the $l = \{x, y\}$-directions, and $\kappa$ is the complex conjugation operator.
\\
When SNN interactions are omitted, the system presents all four symmetries. The addition of any of the SNN interactions removes chiral and particle hole symmetry.
\\
The symmetry change results in a change in topological behavior. In particular, a loss of chiral symmetry results in the well defined corner and edge states in the BDI class of systems changing to a lowered symmetry AI class of systems. The AI system still presents surface states, but they are no longer localized only on the corner or edge but rather a combination of the two.  

\subsection{Invariant}
\label{Section: Invariant}
When a system undergoes a gap closure, the pseudo-momentum in reciprocal space of the system acts as a conserved parameter. This can be exploited to define an invariant which describes the topological phase of the system.   
\\
Two dimensional systems with time-reversal, particle-hole and chiral symmetries have a trivial two dimensional Chern number. However, due to the Hamiltonian of the BDI system being separable, a vectored version of the invariant can be formed from the 1D Chern number (also called the Zak phase) in each independent direction as these values are quantized.\cite{Obana, Liu} This invariant is convenient to use when no SNN interactions are present. 
\\
The Zak phase in its component form ($l \in \{x, y\}$) is described by
\begin{equation}
	\label{Eqn: Zak}
	Z_l (\vec{k}) = -\frac{i}{\pi} \sum^{N_{occ}}_{j = 1} \int^{2 \pi}_{0} \bra{u_j(\vec{k})} \frac{\partial}{\partial k_l}\ket{u_j(\vec{k})}\mathrm{d}k_l, \quad l = \{x, y\} 
\end{equation}
where $j$ runs over all occupied bands, $\ket{u_j(\vec{k})}$ is the periodic Bloch function for the $j^{\mathrm{th}}$ band and $\bra{u_j(\vec{k})} \frac{\partial}{\partial k_l}\ket{u_j(\vec{k})}$ is the Berry connection. This value indicates the topological phase -- $Z_l = 0$ indicates a trivial phase, while $Z_l =2$ indicates a topological phase. 
\\
If the system has inversion symmetry as well, the integral of the Zak phase can be reduced to the contributions at the four time reversal invariant momenta \cite{Fu}, i.e. $\Gamma (\va{k} = \left(0, 0\right))$, $X  (\va{k} = \left(\pi, 0\right))$, $Y   (\va{k} = \left(0, \pi\right))$ and $M  (\va{k} = \left(\pi, \pi\right))$. This can be exploited to define a new invariant using the inversion eigenvalues at each point for the occupied bands. These eigenvalues are calculated using the matrix $\iota H \iota^{-1}$, where $\iota$ is the unitary inversion operator. The inversion eigenvalues can only take on values of $\pm 1$ indicating if the bands have even or odd parity at each point. \cite{ParityInv, ParityInv1} Due to the symmetry present in the system, the product of the inversion eigenvalues at the high symmetry points is always trivial -- the $\Gamma$ and $M$ points always have the same parity eigenvalue. This can be exploited to form a vectored invariant:
\begin{gather}
	\label{eq:Invariant}
	(-1)^{\nu_x}  = \prod_{n_{occ}}\frac{\xi(X)}{\xi(\Gamma)}; \nonumber
	\\
	(-1)^{\nu_y}  = \prod_{n_{occ}} \frac{\xi(Y)}{\xi(\Gamma)}; 
\end{gather}
This invariant is consistent with the vectored Zak phase for the system with no SNN interactions. Since the SNN interactions break chiral and particle-hole symmetry, the Zak phase is no longer well defined for this model. The alternate invariant is defined using only inversion and time-reversal symmetry, so it can still be used to describe the topology of the system, and will therefore be used throughout this work.
\\
Considering only the high symmetry points, the energy and inversion eigenvalues are as follows:
\begin{equation}
	\begin{aligned}
		E_{1(2)}(\Gamma) & =  \pm \left(\delta+ \delta_x+ \delta_{xy}+ \delta_y- \gamma_x- \gamma_x^\prime\right)-\gamma_y- \gamma_y^\prime \\  \xi_{1(2)}(\Gamma)& = \pm 1
		\\
		E_{3(4)}(\Gamma) & = \pm\left(-\delta- \delta_x- \delta_{xy}- \delta_y- \gamma_x- \gamma_x^\prime\right)+\gamma_y+ \gamma_y^\prime, \\ \xi_{3(4)}(\Gamma) & = \mp 1 \\ \\
		E_{1(2)}(X) & = \pm \left(-\delta+ \delta_x+ \delta_{xy}- \delta_y+ \gamma_x- \gamma_x^\prime\right)-\gamma_y- \gamma_y^\prime,  \\ \xi_{1(2)}(X) & = \mp 1
		\\
		E_{3(4)}(X) & = \pm \left(\delta- \delta_x- \delta_{xy}+\delta_y+ \gamma_x- \gamma_x^\prime\right)+\gamma_y+ \gamma_y^\prime, \\ \xi_{3(4)}(X)&= \pm 1
		\\
		\\
		E_{1(2)}(Y) & = \pm \left(-\delta- \delta_x+ \delta_{xy}+ \delta_y- \gamma_x- \gamma_x^\prime\right)+\gamma_y- \gamma_y^\prime, \\ \xi_{1(2)}(Y)&  = \mp 1
		\\
		E_{3(4)}(Y) & = \pm \left(\delta+ \delta_x- \delta_{xy}- \delta_y- \gamma_x- \gamma_x^\prime\right)-\gamma_y+ \gamma_y^\prime, \\ \xi_{3(4)}(Y) &= \pm 1
	\end{aligned}
\end{equation}
Using these expressions, depending on the parameter values, the above expressions could be used to establish the occupied bands and the invariant. A phase transition is possible at any of these high symmetry points, which means that SNN interactions can trigger a phase change.

\section{Numerical Results}
\label{Section: Results}
The band gap and band behavior in reciprocal space gives an indication of where a topological phase transition will occur. A topological phase should correspond to the localization of the probability density of an open system's real space wave functions. 
\\
Using these markers, a system with no SNN interactions will be studied to establish topological phases which warrant the addition of a SNN perturbation. The effect of the SNN interactions on the phases will then be investigated.  

\subsection{2D SSH Model}
\label{Section: SSH results}
The gap conditions for the reciprocal space and real space systems should be similar for the a large enough system due to bulk boundary correspondence. A system size of at least $65 \times 65$ unit cells was the smallest system required to still be representative of the $k$-space gaps. For localization checks in the open boundary case the surface was assumed to be 4 unit cells thick.
\\
The gap size for the periodic system for a given set of parameters was calculated to establish regions in parameter space where topological transitions could be possible. As this is a preliminary system, the smallest degrees of freedom are desired. When studying a purely isotropic model, $\gamma_x = \gamma_y$ and $\gamma_x^\prime =\gamma_y^\prime$, it was found that the system was always metallic. Metallic states are problematic to define topologically as the invariant is affected by the interaction of the bands and changes in their parity. While starting with a metallic phase and adding SNN interactions to cause a non-metallic phase which is well-defined topologically is not problematic, it limits the study. In order to ensure that all cases of the model are well represented, a "nearly isotropic" model is considered, where $\gamma_x = \gamma_y = \gamma$ and $\gamma_x^\prime$ and $\gamma_y^\prime$ are allowed to differ. In order to reduce the parameters, we scale all other parameters by $\gamma$, without loss of generality.  
\\
\begin{figure}[htb]
	\centering
	\includegraphics[width=8.6cm]{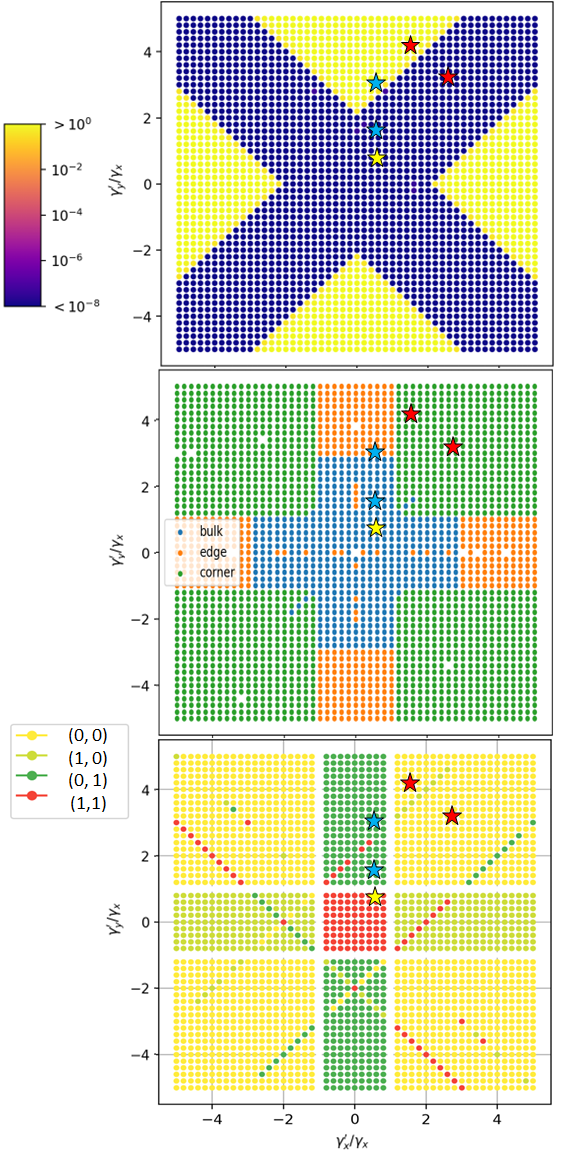}
	\caption{(Top) Bulk band gap, (Middle) valence band maximum surface localization type for a system with $80\times 80$ unit cells with open boundary conditions and (Bottom) parity invariant for a nearly isotropic 2D SSH system. The stars represent the 5 different sample points used to study the different phases possible for the system. }
	\label{Fig: SSH_summ}
\end{figure}
Figure~\ref{Fig: SSH_summ} (Top) indicates that the near isotropic setup can describe both a metallic and non-metallic system. The localization of the non-periodic wave functions of the valence band maximum (VBM) (Figure~\ref{Fig: SSH_summ} (Middle)) shows that the wave functions are localized in the bulk, edge or at the corners in distinct regions of the parameter space. The right panel of the figure shows the corresponding parity invariant, which confirms the analytic results predicted by the discriminant conditions. When $\frac{\gamma_{l}^\prime}{\gamma_l} < 1$, when $l$ is $x$ or $y$, the system is trivial and when $\frac{\gamma_{l}^\prime}{\gamma_l} > 1$, the system is topological. These results show that when the system is topological in both directions, the corner state is preserved despite the metallicity of the system. For a partially trivial system, the metallic system shows a bulk localization, while the non-metallic system shows the corner state predicted by the invariant. This serves as an example of why a metallic system and the localization of its states can be challenging to describe using an invariant quantity. The trivial topological phase always corresponds to a bulk metallic system. 
\\
These results allow the identification of three distinct topological phases -- trivial: $\left(\nu_x, \nu_y\right)$ = $\left(0, 0\right)$, partially trivial: $\left(\nu_x, \nu_y\right)$ = $\left(0, 1\right)$ or $\left(1, 0\right)$ and topological: $\left(\nu_x, \nu_y\right)$ = $\left(1, 1\right)$. These can further be broken down into metallic and non-metallic cases where applicable. These distinctions led to the choice of 5 representative points in parameter space indicated by stars in Figure~\ref{Fig: SSH_summ}. The points consist of a metallic, trivial phase ($M_{00}$) with $\gamma_{x}^\prime = 0.5, \gamma_{y}^\prime = 0.75$ (blue star), metallic and non-metallic, partially trivial phases: $M_{01}$ with $\gamma_{x}^\prime = 0.5$, $NM_{01}$ with $\gamma_{y}^\prime = 1.6$, $\gamma_{x}^\prime = 0.5, \gamma_{y}^\prime = 3$ (yellow stars), and a metallic and non-metallic topological phase: $M_{11}$, with $\gamma_{x}^\prime = 2.5, \gamma_{y}^\prime = 3.2$ and $NM_{11}$ with $\gamma_{x}^\prime = 1.5, \gamma_{y}^\prime = 4.2$ (red stars).

\subsection{2D SSH Model with SNN Interactions}
The addition of SNN interactions breaks chiral and particle-hole symmetry, therefore the corner and edge states are no longer well defined and will instead be labelled as non-bulk states localized on the edge, corner, or a combination of the two. 

\subsubsection{Single Parameter Symmetry Breaking}
\begin{figure*}[htb]
	\centering
	\includegraphics[width=17cm]{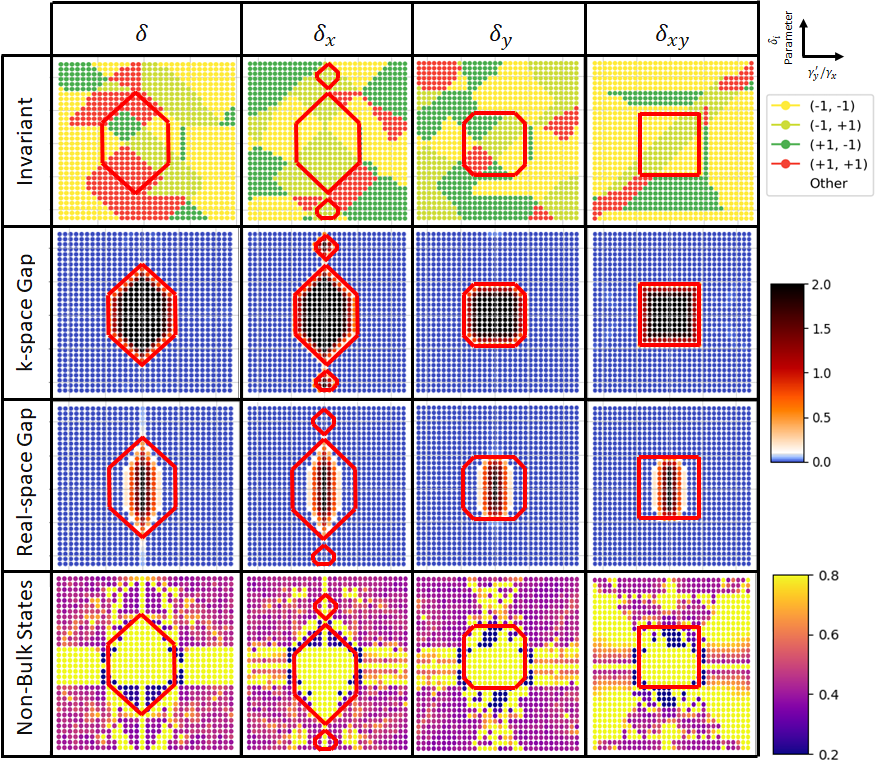}
	\caption{Bulk invariant and gap $E_g$ as well as gap size and VBM state localization for a non-periodic real system with $\gamma_x^\prime = 4.1$ and different SNN parameters as a function of $\gamma_y^\prime$. Parameters are varied in interval $\left[ -5, 5 \right]  $.}
	\label{Fig: cuts}
\end{figure*}
To test the effect of the SNN interactions on the states, they are added to a system with $\gamma_x^\prime = 4.1$, while $\gamma_y^\prime$ and the SNN interaction are allowed to vary. This line is chosen so that a topological and partially topological phase are sampled. The results of the SNN additions are shown in Figure~\ref{Fig: cuts}. $\gamma_y^\prime$ is varied on the horizontal axis, and one SNN parameter is varied on the vertical axis. The other SNN parameters are kept zero respectively.\\
Focusing only on the non-metallic regions shows that the addition of SNN interactions can -- but does not have to -- evoke an immediate change in the invariant, which is consistent with the perturbation causing a symmetry change. In general, the valence band maximum (VBM) state surface localization is consistent with what is expected from the invariant. The states do not always have a one-to-one correlation with the invariant, this is believed to be due to finite size effects.
\\
\begin{figure}[htb]
	\centering
	\includegraphics[width=8.0cm]{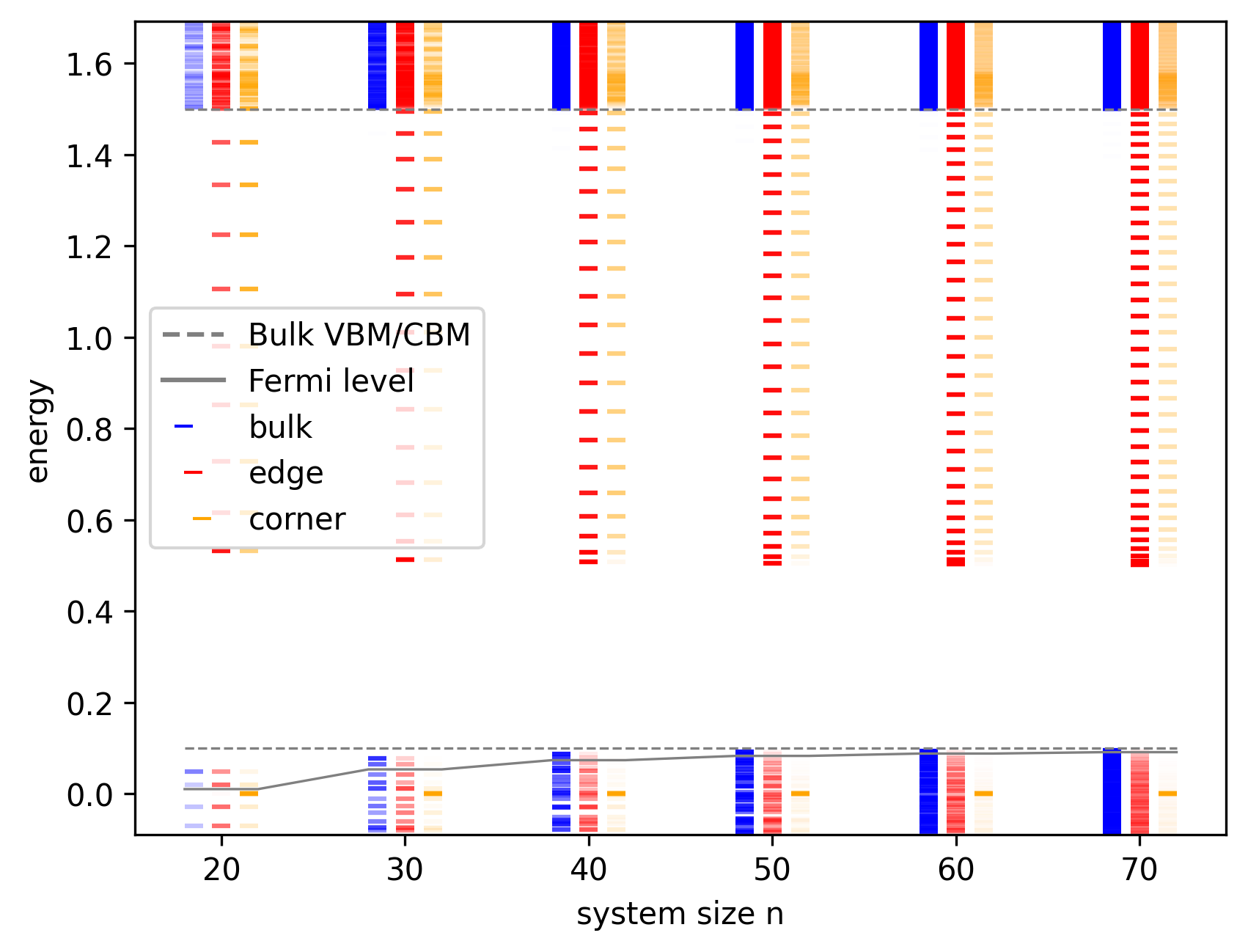} \\
	\includegraphics[width=8.6cm]{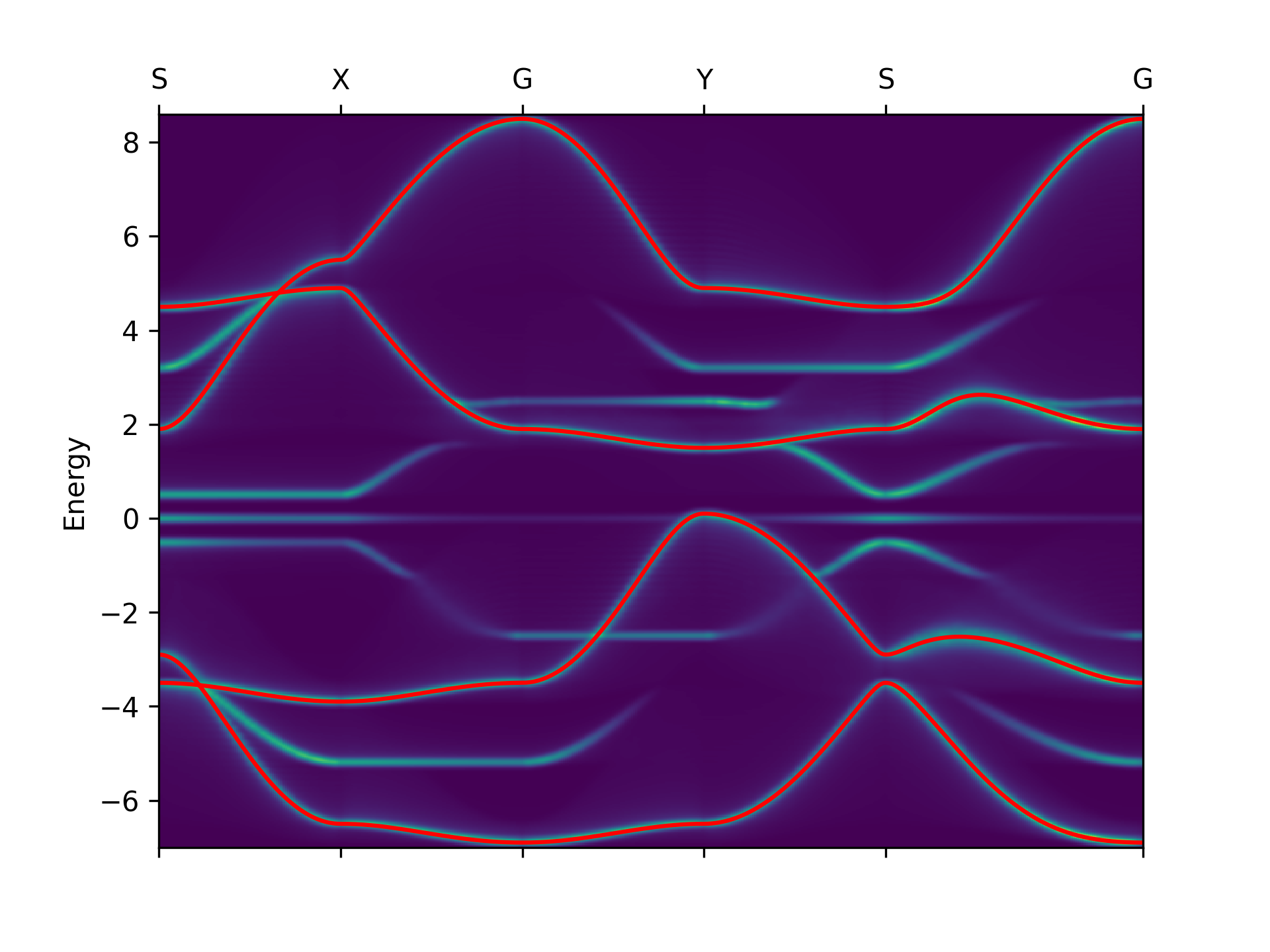} \\
	\includegraphics[width=8.6cm]{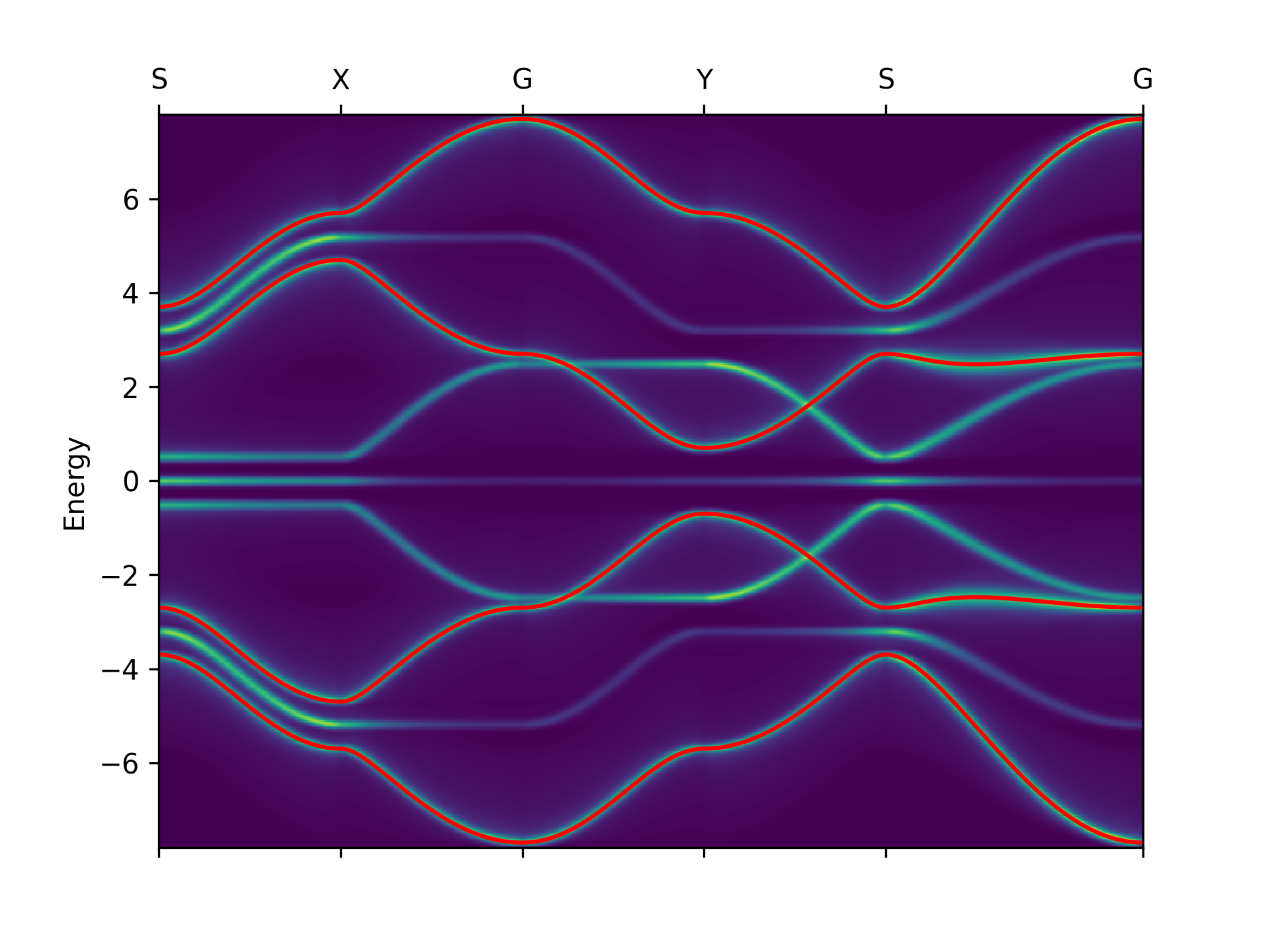}
	\caption{Fully topological non-metallic phase, where $\gamma_x = \gamma_y = 1$, $\gamma_x^\prime = 1.5$ and $\gamma_y^\prime = 4.2$ with $\delta = -0.8$. Top: Eigenvalues and state localization for different system sizes and open boundary conditions. Opacity indicates strength of bulk, edge and corner contributions to the wave functions. Center: Fourier transform of $n=70$ wave functions. Red curves represent the corresponding bulk band structure. Bottom: Reference band structure without SNN interaction.}
	\label{Fig: Z22 states}
\end{figure}
To demonstrate a typical example we choose the topological $NM_{11}$ configuration and a $\delta = -0.8$ SNN interaction. This configuration is non-metallic in the bulk, while the open boundary system is gapless. In Figure~\ref{Fig: Z22 states} (left) the energy eigenvalues and their localization character are shown for several system sizes. The zero energy corner states are still present and unchanged. The SNN term deforms and shifts the bulks though, so that the valence band maximum now sit above these states at $E_{VBM}=0.1$. This can also be seen in the band structure plot in Figure~\ref{Fig: Z22 states} (center). This system also shows low energy edge states in the bulk gap. The corner and edge states are barely affected by the symmetry breaking SNN interaction and retain their energy and dispersion from the SSH reference state, Figure~\ref{Fig: Z22 states} (right). 

\subsubsection{Two Parameter Variations}
With the confirmation that the symmetry breaking caused by a single SNN interaction causes a phase change, a combination of two of the parameters are now introduced for each representative point. To limit the effect on the symmetry of the system, $\delta$-$\delta_{xy}$ and $\delta_x$-$\delta_y$ were studied as pairs so that there are contributions in both directions. \\
Since the study focuses on the changes due to the SNN interactions, the role of symmetry in this process is important to note. When one of the representative points describing the different states is chosen, the system behaves in a qualitatively consistent way in that region irrespective of the exact parameters of the point chosen. Along the diagonal, the high symmetry results in the invariant being ill-defined. 
\\
\begin{figure*}[htb]
	\centering
	\includegraphics[width=17cm]{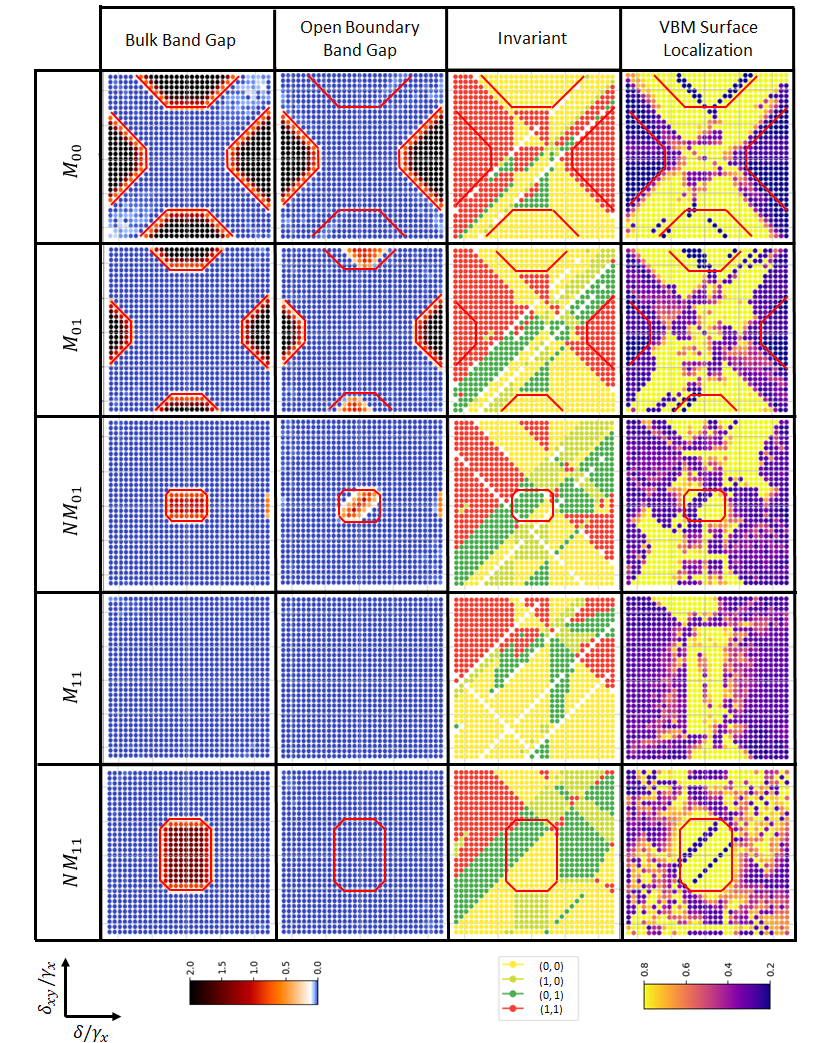}
	\caption{A metallic trivial, metallic and non-metallic partially trivial and a metallic and non-metallic topological system for varying $\delta$ and $\delta_{xy}$.}
	\label{Fig: combo}
\end{figure*} 
For illustrative purposes, we discuss the $\delta$-$\delta_{xy}$ pair. The $\delta_x$-$\delta_{y}$ results are shown in the supplementary material. The first two columns of Figure~\ref{Fig: combo} indicate that, aside from the fully topological metallic state, a gap can be closed or opened by the parameter pair. Both the invariant and surface localization of a state are defined using the valence band maximum. Due to the interaction between the the bands near the Fermi level in a metallic system, different tools beyond the scope of this work are required to study these states. The last two columns of Figure~\ref{Fig: combo} show that the insulating regions are topologically protected. When the region is insulating in the bulk, but only a subset of that region is insulating in the open boundary system, a surface state (confirmed by the invariant) is visible. This exemplifies how the bulk gap closure is necessary, but does not always indicate a topological phase transition. Due to the finite size effect, the VBM surface localization does not fully correspond to the invariant, but would for a larger enough system size. When the symmetry of the Hamiltonian, and by implication the bands is broken, the distinction between edge and corner states is no longer clearly defined. Due to this, we choose to define the topological phases when SNN interactions are included as topologically protected (surface localized) or trivial (bulk localized). 
\\
\begin{figure}[htb]
	\centering
	\includegraphics[width=8.6cm]{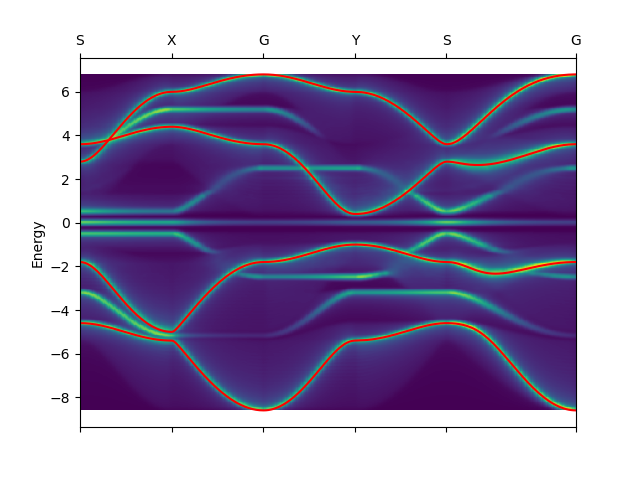} \\
	\includegraphics[width=8.6cm]{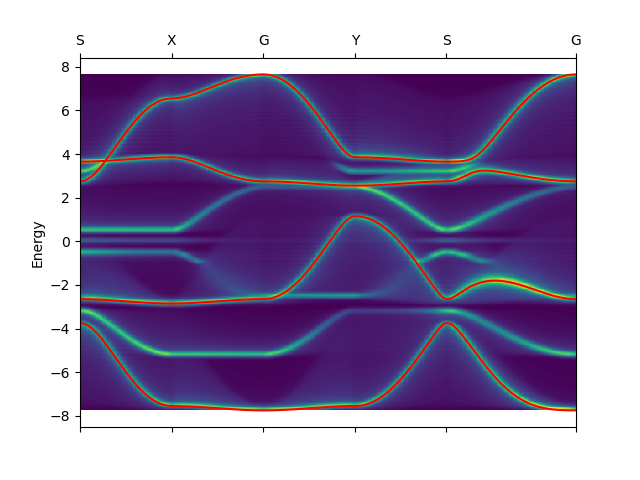} \\
	\caption{Fourier transforms of the open boundary wave functions plotted over reciprocal space, with the bulk wave functions are indicated in red for a $NM_{11}$ system,  where the bulk band gap is less than the open boundary band gap, with (top) $\delta =0.6 $ and $\delta_{xy} = 0.3$ i.e. a fully topological phase, and (bottom) $\delta = -0.9$ and $\delta_{xy} = \pm 0.95$ i.e. a partially topological phase. A system size of $n = 60$ was used.}
	\label{Fig:zero_state_combo}
\end{figure}
In order to further analyze the topologically protected regions, a Fourier transform of the open boundary wave functions was plotted over reciprocal space and compared to the bulk wave functions. In order for a topological surface state to be well defined, it should be in the bulk band gap. Provided, the region where the gap size of the open boundary system is less than or equal to the bulk band gap, the topological state is well defined and in the gap. For example, the left panel of Figure~\ref{Fig:zero_state_combo} shows an example of where the state is in the bulk band gap i.e. $\delta = 0.6$ and $\delta_{xy} = 0.3$ and well defined. This corresponds to a fully topological system where the entire surface state is in the band gap. The right panel of Figure~\ref{Fig:zero_state_combo} with $\delta = -0.9$ and $\delta_{xy} = 0.95$  corresponds to a partially topological phase where only part of the surface state is in the band gap. A trivial phase would have no surface states in the band gap at all.

\section{Conclusion}
\label{Section: Concl}
The addition of second nearest neighbor interaction to a 2D SSH model removes chiral and particle-hole symmetry, changing the topological classification from BDI and AI. For this case a vectorized parity invariant can be defined and the parity eigenvalues for this model are given.

In this model SSN interactions have little or no effect on the surface states but shift and deform the bulk bands. As a result a small perturbation in the form of the SNN interactions can evoke a change allowing a transition from a metallic to a non-metallic system and also from a topological phase to a trivial one or vice versa. The SNN interactions result in band inversions and shifts of the surface states in and out of the band gap.

We find that the non-metallic fully topological phase ($NM_{11}$) is most resilient against the symmetry-breaking perturbation. New topological phases for larger magnitudes of SNN interaction occur in particular for the trivial ($M_{00}$) and partially trivial ($M_{01}$) metallic phases.

\medskip
\textbf{Acknowledgements} \par 
The support of the DSI-NRF Centre of Excellence in Strong Materials (CoE-SM) towards this research is hereby acknowledged. Opinions expressed and conclusions arrived at, are those of the author and are not necessarily to be attributed to the CoE-SM. R.W. also acknowledges support by the Mandelstam Institute for Theoretical Physics.

\medskip

\bibliographystyle{MSP}
\bibliography{References.bib}

\end{document}